\documentclass[showpacs,showkeys,11pt,
preprint,preprintnumbers,nofootinbib,
groupedaddress,superscriptaddress,amsmath,amssymb]{revtex4-1}

\usepackage[dvipdfmx]{graphicx}
\usepackage{graphics}

\usepackage[hypertex]{hyperref}

\newcommand{\dsl}{\hspace{-7pt}/}

\usepackage{amsmath}	
\begin{document}
\title{Reconstruction of Inert Doublet Scalars\\
 at the International Linear Collider}
\preprint{KANAZAWA-13-03}
\preprint{UT-HET-078}

\pacs{12.60.Fr,	 
      13.66.Hk,  
      14.80.Ec,  
      14.80.Fd,  
}
\keywords{Extended Higgs Theory, Electron Positron Colliders}
\author{Mayumi Aoki}
\email{mayumi@hep.s.kanazawa-u.ac.jp}
\affiliation{Institute for Theoretical Physics, Kanazawa University,
Kanazawa 920-1192, Japan}
\affiliation{Max-Planck-Institut f\"ur Kernphysik,
Saupfercheckweg 1, 69117 Heidelberg, Germany}
\author{Shinya Kanemura}
\email{kanemu@sci.u-toyama.ac.jp}
\affiliation{Department of Physics, University of Toyama, Toyama
930-8555, Japan}
\author{Hiroshi Yokoya}
\email{hyokoya@sci.u-toyama.ac.jp}
\affiliation{Department of Physics, University of Toyama, Toyama
930-8555, Japan}
\date{\today}

\begin{abstract}
We study collider signatures for extra scalar bosons in the inert
 doublet model at the international linear collider (ILC).
The inert doublet model is a simple extension of the standard model by 
 introducing an additional isospin-doublet scalar field which is odd
 under an unbroken $Z_2$ symmetry. 
The model predicts four kinds of $Z_2$-odd scalar bosons, and the
 lightest of them becomes stable and a candidate of the dark matter as
 long as it is electrically neutral. 
Taking into account the constraints from various theoretical and
 phenomenological conditions, we perform a simulation study for the
 distinctive signatures of the extra scalars over the standard-model
 background contributions at the ILC with the center-of-mass energy of
 $\sqrt{s}=250$~GeV and 500~GeV.
We further discuss observables for determination of the mass of the
 scalars. 
We find that the parameter regions which cannot be detected at the large
 hadron collider can be probed at the ILC.
\end{abstract}
\maketitle

\section{Introduction}

In July 2012, a Higgs-like particle was found at the large hadron
collider (LHC)~\cite{Aad:2012tfa,Chatrchyan:2012ufa}. 
The particle looks like mostly a Higgs boson in the standard model (SM),
but the detail properties of the particle and the whole structure of the
Higgs sector have not yet been revealed.
It is widely believed that the SM has to be extended, since it cannot
explain the dark matter, (tiny) neutrino masses, and the baryon
asymmetry in the Universe, etc. 
Although in the Higgs sector of the SM, only one ${\rm
SU}(2)_{L}$-doublet scalar field is introduced, there is no theoretical
guideline for this choice. 
Thus, the Higgs sector may be a solid target to probe new physics beyond
the SM.

The inert doublet model (IDM) is one of the simplest extensions of the
SM, where an additional ${\rm SU}(2)_{L}$-doublet scalar field is
introduced, which is odd under the unbroken $Z_2$
symmetry~\cite{Barbieri:2006dq,Deshpande:1977rw}. 
As in the case in the general two Higgs doublet model, four kinds of
additional scalars appear as physical states, namely neutral
$CP$-even state ($H$), neutral $CP$-odd state ($A$) and charged scalar
states ($H^{\pm}$), all of which are called inert scalars. 
Due to the $Z_2$ symmetry, Yukawa interactions of the inert scalars to
SM fermions are forbidden, and the possible flavor-changing neutral
current is absent at the tree level.
The lightest inert particle (LIP) is stable, because of the
$Z_2$-parity conservation. 
Therefore, the model provides a dark matter
candidate~\cite{Barbieri:2006dq,LopezHonorez:2006gr,Gustafsson:2007pc,%
Dolle:2009fn,Honorez:2010re,LopezHonorez:2010tb,Gustafsson:2012aj,%
Goudelis:2013uca}.
In addition to that, the model has rich phenomenological features such
as the electroweak symmetry breaking~\cite{Hambye:2007vf},
electroweak phase transition~\cite{Chowdhury:2011ga,Borah:2012pu,%
Gil:2012ya,Cline:2013bln}, 
radiatively generating neutrino masses by introducing $Z_2$-odd
right-handed neutrinos~\cite{Ma:2006km}, and
leptogenesis~\cite{Ma:2006fn,Suematsu:2009ww}, etc. 

Collider signatures of the inert scalars in the IDM have been studied in
the literature~\cite{Barbieri:2006dq,Cao:2007rm,Lundstrom:2008ai,%
Dolle:2009ft,Miao:2010rg,Gustafsson:2012aj,Goudelis:2013uca}.
In Ref.~\cite{Lundstrom:2008ai}, bounds on the inert scalar masses are
obtained by using the experimental results at the LEP
II~\cite{Acciarri:1999km,Abbiendi:1999ar,Abdallah:2003xe}. 
Since the inert scalars do not have QCD interactions, it is not suited
to search for them at hadron colliders. 
Even though the parameter regions where the inert scalars could be
discovered at the
LHC are pointed out~\cite{Dolle:2009ft,Miao:2010rg,Gustafsson:2012aj},
detailed analysis on these scalars such as the precise 
determination of these masses and quantum numbers would be difficult. 

In this letter, we study collider phenomenology for the inert scalars
at the international linear collider (ILC).
As it is a machinery for precision measurements of Higgs boson
properties, the extended Higgs sector can also be investigated in
details. 
We study the characteristic signatures, corresponding backgrounds and
the kinematical observables in the processes of $HA$ associated
production as well as $H^+H^-$ pair production. 
Earlier studies can be found e.g.\ in
Refs.~\cite{Aoki:2010tf,Asano:2011aj} where the charged scalar pair 
production process is studied in detail, and in
Ref.~\cite{Kanemura:2012ha} where $HA$ associated production is also
studied. 
On the other hand, our study includes all the available processes and
decay modes, and the simulation analysis for the signal and the
background contributions with appropriate kinematical cuts.
Furthermore, we also discuss the procedure for the mass determination of
the inert scalars which can be performed at an early stage of the 
experiment, and introduce a new observable which could be useful to
determine the inert scalar masses more precisely than the variables used
in Refs.~\cite{Asano:2011aj,Kanemura:2012ha}. 

The rest of the letter is organized as follows.
In Sec.~\ref{sec:idm}, we review the inert doublet model and introduce 
the benchmark points used in our simulation study.
Then we present the simulation studies for observing the inert scalars
and determining the masses of them in Sec.~\ref{sec:sim}. 
Sec.~\ref{sec:dis} and Sec.~\ref{sec:conc} are devoted to discussions
and conclusions, respectively.
In Appendix, we evaluate a new observable at $e^+e^-$ colliders which
is used in our analysis for the mass determination.

\section{The Inert Doublet Model}\label{sec:idm}

In the IDM, the two scalar doublet fields, $\Phi_1$ and $\Phi_2$, have
even and odd parity under the $Z_2$ symmetry, respectively. 
The most general scalar potential can be written as 
\begin{align}
 V(\Phi_1,\Phi_2) &=
 \mu_1^2\left|\Phi_1\right|^2 + \mu_2^2\left|\Phi_2\right|^2
 + \frac{\lambda_1}{2}\left|\Phi_1\right|^4
 + \frac{\lambda_2}{2}\left|\Phi_2\right|^4
 + \lambda_3\left|\Phi_1\right|^2\left|\Phi_2\right|^2 \nonumber \\
 & + \lambda_4\left|\Phi_1^\dagger\Phi_2\right|^2
 + \left\{\frac{\lambda_5}{2}\left(\Phi_1^\dagger\Phi_2\right)^2
 + {\rm H.c.}\right\},
\end{align}
with seven real parameters $(\mu_1^2,\mu_2^2,\lambda_{1},
\lambda_{2},\lambda_{3},\lambda_{4},\lambda_5)$.
We note that the potential is invariant under the $CP$ transformation.
The potential has to satisfy theoretical constraints, such as the vacuum
stability~\cite{Deshpande:1977rw,Ginzburg:2010wa} and the 
perturbativity~\cite{Barbieri:2006dq}. 
By the vacuum stability at the tree level, the quartic terms are
constrained as $\lambda_1>0$, $\lambda_2>0$,
$\sqrt{\lambda_1\lambda_2}+\lambda_3>0$, and
$\sqrt{\lambda_1\lambda_2}+\lambda_3+\lambda_4-|\lambda_5|>0$~\cite{Deshpande:1977rw}. 
We consider the case where $\mu_1^2<0$,
$\lambda_1\mu_2^2>\lambda_3\mu_1^2$ and 
$\lambda_1\mu_2^2>(\lambda_3+\lambda_4+|\lambda_5|)\mu_1^2$ are
satisfied~\cite{Deshpande:1977rw}, so that $\Phi_2$ does not acquire the
vacuum expectation value (VEV)~\cite{Fayet:1974fj} and only $\Phi_1$
plays a role of the ``Higgs-boson''. 
By denoting
\begin{align}
 \Phi_1= \left(\begin{array}{c}
	   0 \\ \frac{1}{\sqrt{2}}(v+h)
		\end{array}\right), \quad
 \Phi_2= \left(\begin{array}{c}
	   H^+ \\ \frac{1}{\sqrt{2}}(H+iA)
		\end{array}\right), 
\end{align}
where $v$ is the VEV, $v=\sqrt{-2\mu_1^2/\lambda_1}(\simeq246$~GeV), the
masses of these scalars are expressed as 
\begin{subequations}
\begin{align}
 &m^2_h = \lambda_1 v^2,\\
 &m^2_{H^+}=\mu_2^2+\frac{1}{2}\lambda_3 v^2,\\
 &m^2_{H}=\mu_2^2+\frac{1}{2}(\lambda_3+\lambda_4+\lambda_5)v^2,\\
 &m^2_{A}=\mu_2^2+\frac{1}{2}(\lambda_3+\lambda_4-\lambda_5)v^2.
\end{align}
\end{subequations}
Thus, the seven parameters in the Higgs potential can be replaced by
the VEV $v$, four masses of the Higgs boson and inert scalars,
$(m_h,m_{H^+},m_{H},m_{A})$, the scalar self-coupling constant
$\lambda_2$, and $\lambda_H(\equiv\lambda_3+\lambda_4+\lambda_5)$ for
example.
To force the LIP to be electrically neutral, so that it can be a
candidate of the dark matter, $\lambda_4<|\lambda_5|$ must be
satisfied~\cite{Ginzburg:2010wa}. 
Depending on the sign of $\lambda_5$, either $H$ or $A$ becomes the
LIP. 
Since phenomenological constraints and collider signatures are
exchangeable between the two cases, we take $H$ as the LIP
($\lambda_5<0$) hereafter. 

In this letter, we consider four benchmark points for the masses of
inert scalars listed in Table~\ref{tab:ilc}, which satisfy all the 
available theoretical and also phenomenological
constraints~\cite{Gustafsson:2010zz}.
The bounds on the masses of inert scalars are briefly summarized as
follows.
By the constraints from dark matter relic abundance and direct searches,
the mass of LIP should be $40\lesssim
m_H\lesssim80$~GeV~\cite{LopezHonorez:2006gr,Dolle:2009fn,%
Gustafsson:2012aj}~\footnote{%
There exists a case in which all the scalar masses are heavy ($>500$~GeV)
and degenerated~\cite{LopezHonorez:2006gr}.
However, since collider searches at the ILC are difficult, we don't
consider such case in this study.
}. 
For $m_H<80$~GeV, the mass of the second neutral scalar has to satisfy
$|m_A-m_H|<8$~GeV or $m_A>100$~GeV, to avoid the bounds from the direct
searches at the LEP experiments~\cite{Lundstrom:2008ai}. 
The mass differences of inert scalars result into additional
contributions~\cite{Barbieri:2006dq} to the electroweak $S$ and $T$ 
parameters.
To be consistent with the current experimental
data~\cite{Beringer:1900zz}, 
\begin{align}
 \Delta T\simeq 1.08\left(\frac{m_{H^{\pm}}-m_{H}}{v}\right)
 \left(\frac{m_{H^{\pm}}-m_{A}}{v}\right)=0.07\pm0.08
\end{align}
must be satisfied.
In Ref.~\cite{Gustafsson:2012aj} these constraints are more extensively
studied.

In the four benchmark points, the mass of $H$ is fixed to 65~GeV, so
that it does not induce the invisible decay of the SM Higgs boson.
While it could be up to $\sim80$~GeV concerning the dark matter relic
abundance
analysis~\cite{LopezHonorez:2006gr,Dolle:2009fn,Honorez:2010re,%
LopezHonorez:2010tb,Gustafsson:2012aj,Goudelis:2013uca}, the
collider phenomenology does not change qualitatively by varying it in
that region. 
Above that value, $HH\to W^+W^-$ process gives a too large annihilation
cross section, so that the predicted relic density in the IDM results
into a lower value than that from the WMAP experiment.
Below 80~GeV, the processes $HH\to f\bar{f}$ can give proper
annihilation cross sections using the Higg boson mediated
diagram, at the same time with avoiding the constraints from direct
searches by choosing the $hHH$ coupling constant $\lambda_H$ as an
appropriate value.
Here, we have assumed the mass of the Higgs boson to $m_h=126$~GeV. 
When the LIP is below 63GeV, the collider signatures would not change
qualitatively as well, however, the discovery channel of the new physics
signal would be the invisible decay of the Higgs-boson at the LHC. 
We study the case where the masses of $H$ and $A$ are close to each
other (I, III) for which the LEP and LHC experiments can not
probe~\cite{Lundstrom:2008ai,Cao:2007rm,Dolle:2009ft,Miao:2010rg,%
Gustafsson:2012aj}.
The other cases are when $m_A-m_H$ is medium (II) or large such that
the $Z$-boson from $A\to HZ$ becomes on-shell (IV).
For the $W$-bosons in $H^\pm\to W^\pm H$, we consider the
off-shell (I, II) and on-shell (III, IV) cases. 

$\lambda_2$ and $\lambda_H$ do not enter in our collider analysis. 
$\lambda_H$ can be determined by the relic abundance which should be
consistent with the recent WMAP search result~\cite{LopezHonorez:2006gr}
and also the bounds on the cross section of dark matter direct
production.
On the other hand, the method to determine $\lambda_2$ has not yet been
established.

For our four benchmark points, the production cross sections of inert
scalars at the ILC are large enough to be tested.
In Table~\ref{tab:ilc}, we also list the cross sections of $HA$
production and $H^+H^-$ production at $\sqrt{s}=250$~GeV
and 500~GeV.
The cross section of $HA$ production takes the largest value, i.e.\
186~fb at $\sqrt{s}=190$~GeV, 78~fb at $\sqrt{s}=280$~GeV, and 46~fb at 
$\sqrt{s}=350$~GeV in the cases (I, III), (II), and (IV), respectively. 
The cross section of $H^+H^-$ production takes the largest value, i.e.\
96~fb at $\sqrt{s}=380$~GeV and 53~fb at $\sqrt{s}=500$~GeV for the
cases (I, II) and (III, IV), respectively.
At $\sqrt{s}=1$~TeV, the production cross sections are about 10~fb and
20~fb for $HA$ production and $H^+H^-$ production, respectively, for all
the four benchmark points. 
\begin{table}[b]
 \begin{tabular}{c||ccc|cc}
  & \multicolumn{3}{c|}{Inert scalar masses} &
  \multicolumn{2}{c}{ILC cross sections [$\sqrt{s}=250$~GeV (500~GeV)]} \\
  & $m_{H}$~[GeV] & $m_{A}$~[GeV] & $m_{H^\pm}$~[GeV] &
  $\sigma_{e^+e^-\to HA}$~[fb] & $\sigma_{e^+e^-\to H^+H^-}$~[fb] \\
  \hline
  (I)   & 65. & 73.  & 120. & 152. (47.) & 11. (79.) \\
  (II)  & 65. & 120. & 120. & 74. (41.)  & 11. (79.) \\
  (III) & 65. & 73.  & 160. & 152. (47.) & 0. (53.) \\
  (IV)  & 65. & 160. & 160. & 17. (35.)  & 0. (53.)
 \end{tabular}
 \caption{Masses of inert scalars and ILC cross sections
 for our four benchmark points.}\label{tab:ilc}
\end{table}
For the cases (II, IV), $H^{\pm}$ decays into $W^\pm H$ predominantly,
where we admit the $W$-boson to be off-shell if $m_{H^{\pm}}-m_{H}<m_W$. 
While for the cases (I) and (III), $H^{\pm}\to W^\pm A$ decay would be
sizable as well, with the branching ratios about 32\% and 27\%,
respectively.
The decay of the $A$-boson is dominated by $A\to Z^{(*)}H$.

There are two important effects in the property of the SM-like
Higgs boson, the invisible decay~\cite{Cao:2007rm,Swiezewska:2012eh} and
the charged scalar loop contribution to the two-photon decay
amplitude~\cite{Arhrib:2012ia,Swiezewska:2012eh,Goudelis:2013uca}. 
The invisible decay mode opens if $m_h>2m_H$. 
If this is the case, the branching ratio for the invisible decay mode 
can be typically several tens
percent~\cite{Cao:2007rm,Swiezewska:2012eh}. 
Thus, it could be discovered at the LHC~\cite{Bai:2011wz}.
The two-photon branching ratio of the SM-like Higgs boson can be
directly related to $R_{\gamma\gamma}=\sigma(pp\to 
h\to\gamma\gamma)/\sigma(pp\to h\to\gamma\gamma)_{\rm SM}$, because the
production cross-section of the Higgs-boson at the LHC is not modified
in the IDM. 
It is shown~\cite{Swiezewska:2012eh} that $R_{\gamma\gamma}$ can be
enhanced with relatively light $H^{\pm}$ ($m_{H^\pm}\lesssim 130$~GeV),
negative $\lambda_3$ and no invisible decays.

\section{Collider signatures in the IDM at the ILC}\label{sec:sim}

In this section, we perform the simulation studies for the detection and
the mass determination of the inert scalars in the IDM at the ILC. 
The kinematical distributions are calculated by using {\tt
MadGraph}~\cite{Alwall:2011uj} at the parton level with basic cuts for
event generations. 
For charged leptons, we set $p_{T}^{\ell}>1$~GeV, $|\eta_{\ell}|<2.5$
and $\Delta R_{\ell\ell}>0.2$ for the isolation requirement, where $p_T$
is the transverse momentum, $\eta$ is the pseudo-rapidity, and $\Delta 
R (=\sqrt{\Delta\eta^2+\Delta \phi^2})$ is the distance of the two
particle in the $\eta-\phi$ plane. 
For jets, we restrict ourselves with $p_{T}^{j}>5$~GeV,
$|\eta_j|<2.5$ and $\Delta R_{jj}, \Delta R_{\ell j}>0.4$. 
Furthermore, we require the missing transverse energy
$E\dsl_{T}$ to be greater than 10~GeV to reduce background events from
two photon scattering processes.
The two photon scattering processes and QED radiation effects are not
estimated in our analysis.

We note that, these cuts may be rather conservative, so that our
parton level analysis makes sense.
At $e^+e^-$ colliders, hadronic final-states would be utilized even if
they do not form narrow jets.
Then, the cuts on $p_T$ and the isolation for partons may not be
necessary. 
We include these cuts so that the number of jets in an event can be more
easily accounted the number of outgoing partons in the processes. 
In case we loose these cuts, the number of available events would
increase, but we expect more background contributions from the events
with less partons. 
An isolation requirements for leptons may be weakened as well in the
real experiment.

\subsection{$e^+e^-\to HA$ process}

First, we consider the $HA$ production process followed by $A\to Z^{(*)}H
$ decay. 
Since $H$ is neutral and stable, it escapes from detection.
Thus, it gives the signature with a dilepton (dijet) plus large missing
energy. 
Expected background contributions come from dilepton (dijet) plus two
neutrinos production in the SM.

First we study the collider signature of these events at
$\sqrt{s}=250$~GeV, for the parameter sets (I, III) and (II) where
$A$ decays into $H$ and off-shell $Z$-boson. 
To reduce the SM background contributions, we apply following
kinematical cuts; the scaled acoplanarity $\bar\phi_{\rm acop}$, which
is the acoplanarity\footnote{%
The acoplanarity angle $\phi_{\rm acop}$ is defined as the supplement of
the difference of azimuthal angles of the leptons (jets).}
multiplied by the sine of the smallest angle between a lepton (jet)
and the beam axis, is larger than $100^{\circ}$; 
$|\cos\theta_{\ell\ell}|<0.8$ for dilepton or $|\cos\theta_{jj}|<0.6$
for dijet, where $\theta_{\ell\ell(jj)}$ is the polar angle of the
dilepton (dijet) 3-momenta.

\begin{figure}[t]
 \begin{center}
  \includegraphics[width=.33\textwidth,clip]{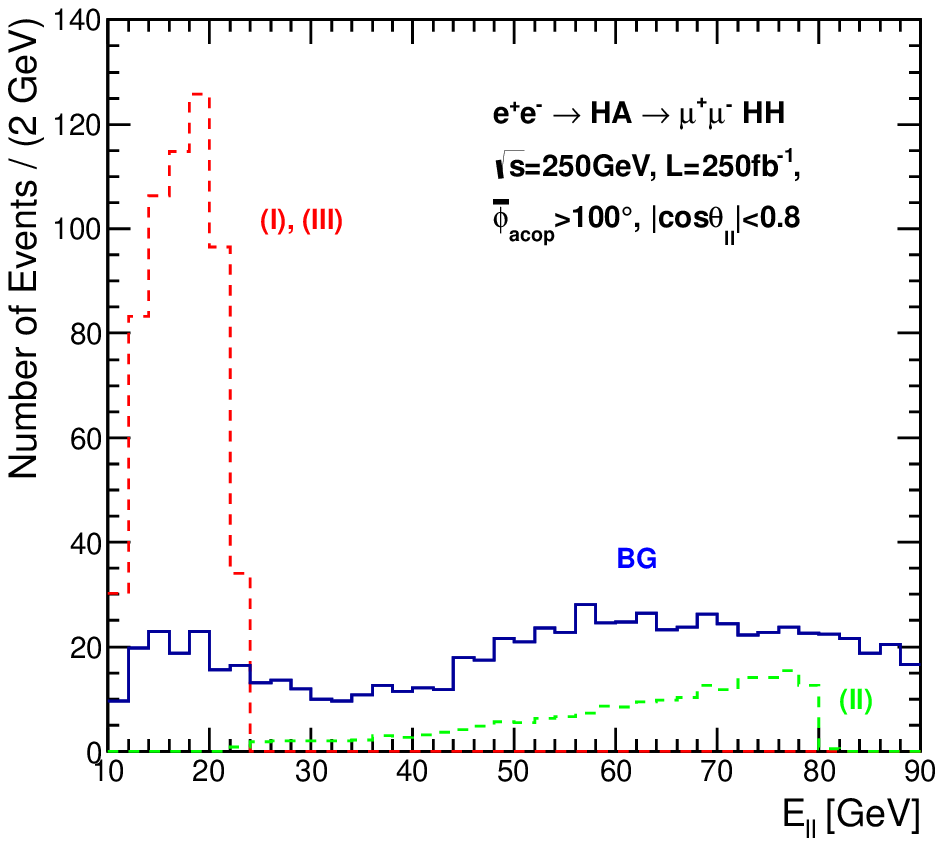}
  \includegraphics[width=.33\textwidth,clip]{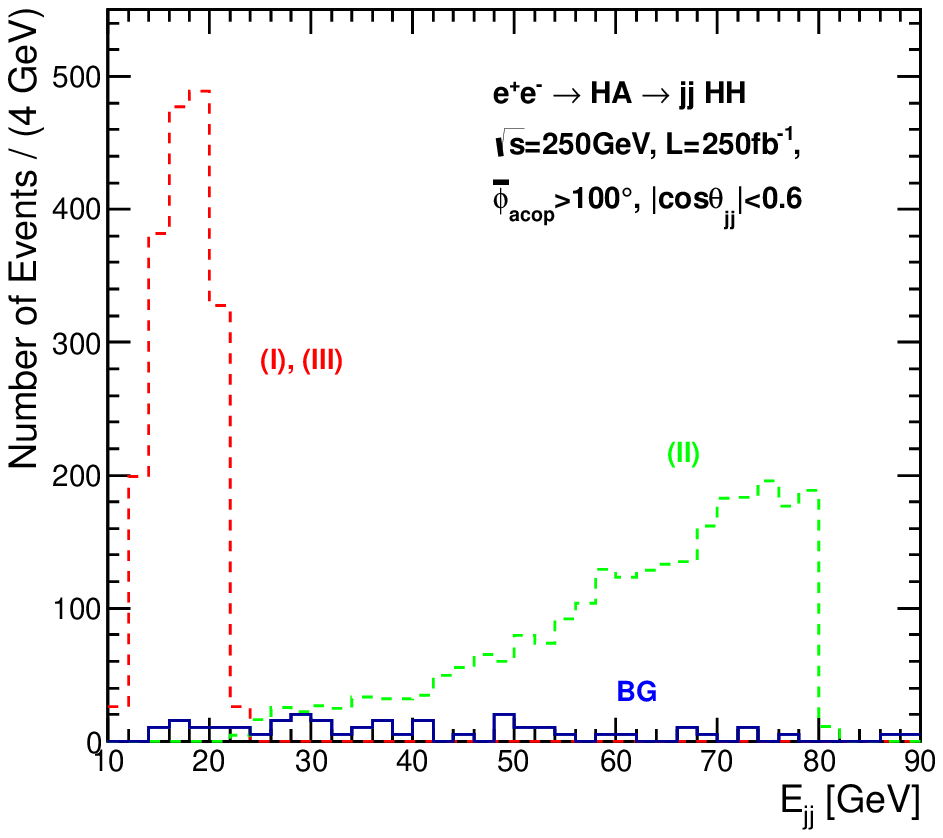} \\
  \includegraphics[width=.33\textwidth,clip]{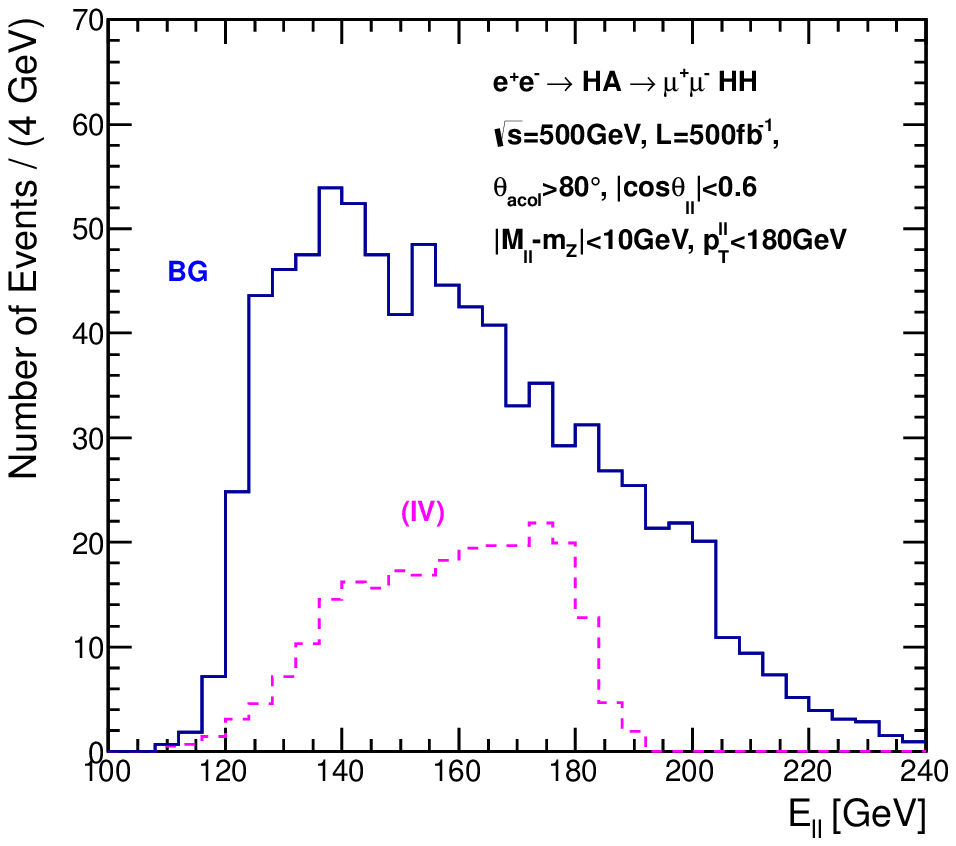}
  \includegraphics[width=.33\textwidth,clip]{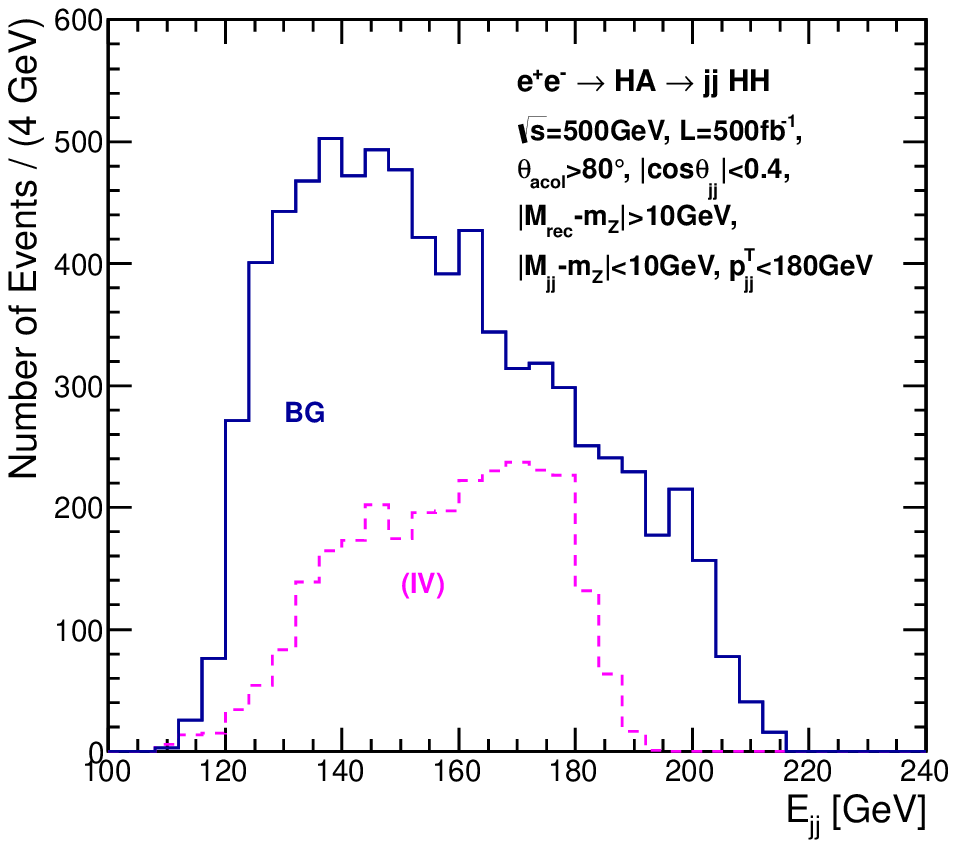}
  \caption{Distributions of dilepton (dijet) energy in dilepton (dijet)
  plus missing energy event at the ILC with $\sqrt{s}=250$~GeV and ${\cal
  L_{\rm int}}=250$~fb$^{-1}$ (top), and with $\sqrt{s}=500$~GeV and
  ${\cal L_{\rm int}}=500$~fb$^{-1}$ (bottom).}
  \label{fig:HA_Kine}
 \end{center}
\end{figure}

In the top left (right) panel in Fig.~\ref{fig:HA_Kine}, we see the
dilepton (dijet) energy distributions for the signal and background
processes at $\sqrt{s}=250$~GeV with the integrated luminosity of ${\cal
L_{\rm int}}=250$~fb$^{-1}$. 
After the cuts described above, background events are well reduced. 
The difference of the number of the signal events in dilepton and dijet
signature comes from the branching ratio of the $Z$-boson, and that of
the background events comes from the absence of $WW$ production in the
dijet case. 

The endpoints of the $E_{\ell\ell(jj)}$ distribution are related with
the masses of $H$ and $A$ as
\begin{align}
 E_{\ell\ell(jj)}^{\rm max/min} =\frac{\sqrt{s}}{4}
 \left(1-\frac{m_H^2}{m_A^2}\right)
 \left[1-\frac{m_H^2}{s}+\frac{m_A^2}{s}
 \pm\lambda(1,\frac{m_H^2}{s},\frac{m_A^2}{s})\right],\label{eq:endoff}
\end{align}
where $\lambda(a,b,c)=\sqrt{a^2+b^2+c^2-2(ab+bc+ca)}$.
Thus, the masses can be determined by measuring these endpoints.
Since the distribution is quite steep around the maximum value,
$E_{\ell\ell(jj)}^{\rm max}$ should be a good observable to be measured
precisely.
It becomes $24$~GeV even for the cases with small mass splitting (I,
III), and $80$~GeV for the case (II) at $\sqrt{s}=250$~GeV. 
On the other hand, $E_{\ell\ell(jj)}^{\rm min}$ measurement may be
difficult since the distribution is gradual around the minimum, and
$E_{\ell\ell(jj)}^{\rm min}$ is too small for the cases
(I, III)~[$E_{\ell\ell(jj)}^{\rm min}=2.4$~GeV at $\sqrt{s}=250$~GeV]. 

In the bottom left (right) panel in Fig.~\ref{fig:HA_Kine}, we show the
$E_{\ell\ell(jj)}$ distributions for the parameter set (IV) with
$\sqrt{s}=500$~GeV and ${\cal L_{\rm int}}=500$~fb$^{-1}$.
To reduce SM background contributions, kinematical cuts are applied;
$\theta_{\rm acol}>80^{\circ}$ where the acollinearity angle
$\theta_{\rm acol}$ is defined as the supplement of the opening angle of
the dilepton (dijet); $|M_{\ell\ell(jj)}-m_{Z}|<10$~GeV;
$p_{T}^{\ell\ell(jj)}<180$~GeV; $|\cos\theta_{\ell\ell}|<0.6$ for the
dilepton case, or $|\cos\theta_{jj}|<0.4$ and $|M_{\rm
rec}-m_{Z}|>10$~GeV for the dijet case, where $M_{\rm rec}$ is the
recoil mass defined as the invariant mass of the missing 4-momenta. 
In the case with the on-shell $Z$-boson, information of the masses can
be obtained from the endpoints of the $E_{\ell\ell(jj)}$ distribution as 
\begin{align}
 E_{\ell\ell(jj)}^{\rm max/min}
 =\gamma_{A}\hat{E}\pm\beta_A\gamma_{A}\hat{p},\label{eq:endon}
\end{align}
where $\hat{E}=(m_{A}^2-m_{H}^2+m_Z^2)/(2m_{A})$,
$\hat{p}=m_{A}/2\cdot\lambda(1,m_{H}^2/m_{A}^2,m_Z^2/m_{A}^2)$,
$\gamma_A=(s-m_{H}^2+m_{A}^2)/(2\sqrt{s}m_{A})$ and
$\beta_A\gamma_A=\sqrt{s}/(2m_{A})\cdot\lambda(1,m_{H}^2/s,m_{A}^2/s)$.
In spite of the finite-width effect of the $Z$-boson, the endpoints
would be seen as sharp edges of the signal plateau.
For the case (IV), the maximum and minimum of the dilepton (dijet)
energies are $E_{\ell\ell(jj)}^{\rm min}=134$~GeV and
$E_{\ell\ell(jj)}^{\rm max}=181$~GeV at $\sqrt{s}=500$~GeV. 
The signal and background distributions are similar for both the
dilepton and dijet cases, while the expected number of events are 10
times larger for the dijet case than that for the dilepton case. 
Thus, the statistical errors are small in the hadronic signature.
On the other hand, the systematical errors would be negligible in the
dilepton case due to the fine resolution of the lepton momentum
measurement. 

For the case with the off-shell $Z$-boson, from the maximum of the
dilepton (dijet) invariant-mass distribution, the difference of the two
scalar masses can be determined as
\begin{align}
M_{\ell\ell(jj)}^{\rm max} = m_{A}-m_{H}.\label{eq:mmax}
\end{align}
We note that, for the case with small mass splitting, the measurement
would be affected by QED radiation backgrounds and acceptance cuts.

\subsection{$e^+e^-\to H^+H^-$ process}

We here turn to the $H^+H^-$ pair production, where $H^\pm$
predominantly decays into $HW^{\pm}$, and $W^{\pm}$ further into 
$\ell^{\pm}\nu$ or $q\bar{q}'$. 
We study the semi-leptonic and all-hadronic decay modes as successful
signatures. 

First, we study the semi-leptonic decay mode, where the signature is a
charged lepton plus dijet plus large missing energy. 
The expected leading background process is $\tau^\pm\nu jj$ production
followed by the leptonic decay of $\tau$. 
The $\ell^\pm\nu jj$ background process can be reduced by requiring
a large recoil mass in the event.
The contribution from production of $\mu^+\mu^-jj$ and missing
particles, where one of the muons goes out of the acceptance region,
also turns out to be negligible. 
The event simulation for the case (I) [the case (III)] is the same as
that for the case (II) [the case (IV)], but the overall normalization is
multiplied by about a half, which is the square of the $H^\pm\to W^\pm H$
branching ratio, 68\% [73\%].

\begin{figure}[t]
 \begin{center}
  \includegraphics[width=0.328\textwidth,clip]{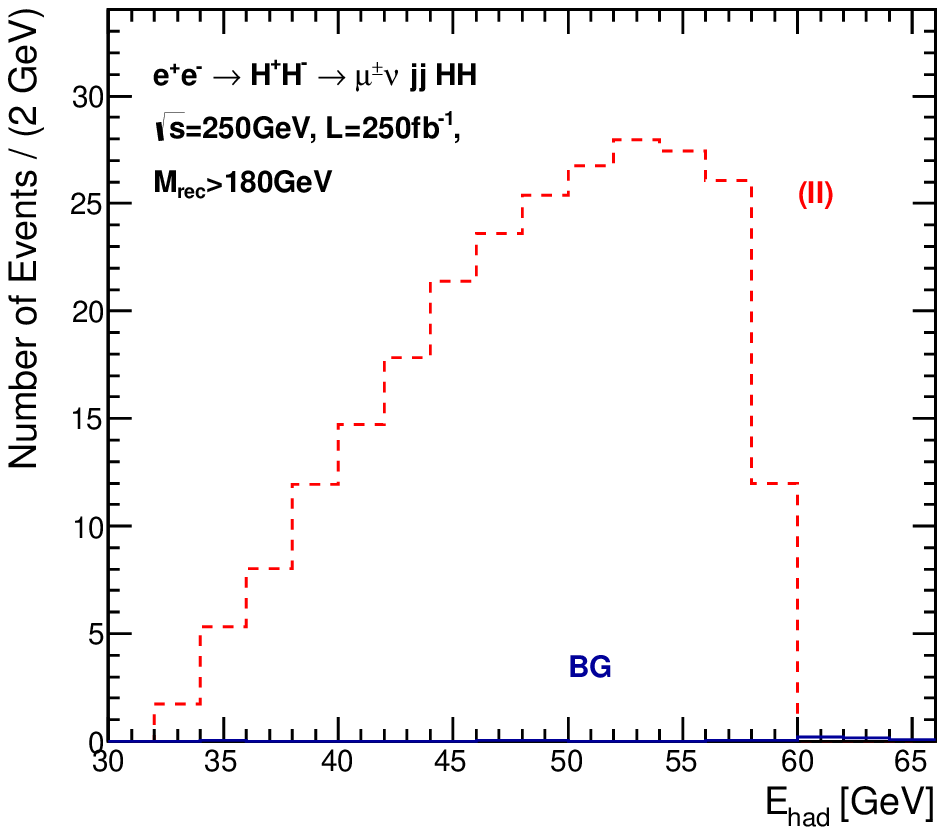}
  \includegraphics[width=0.328\textwidth,clip]{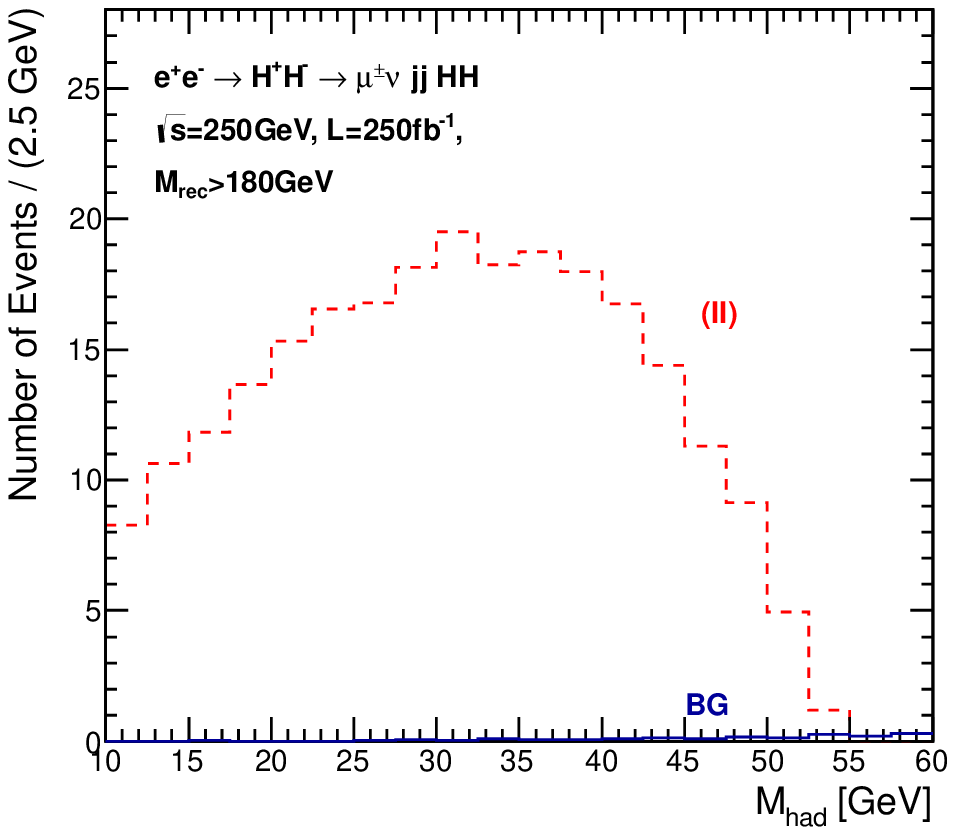}
  \includegraphics[width=0.328\textwidth,clip]{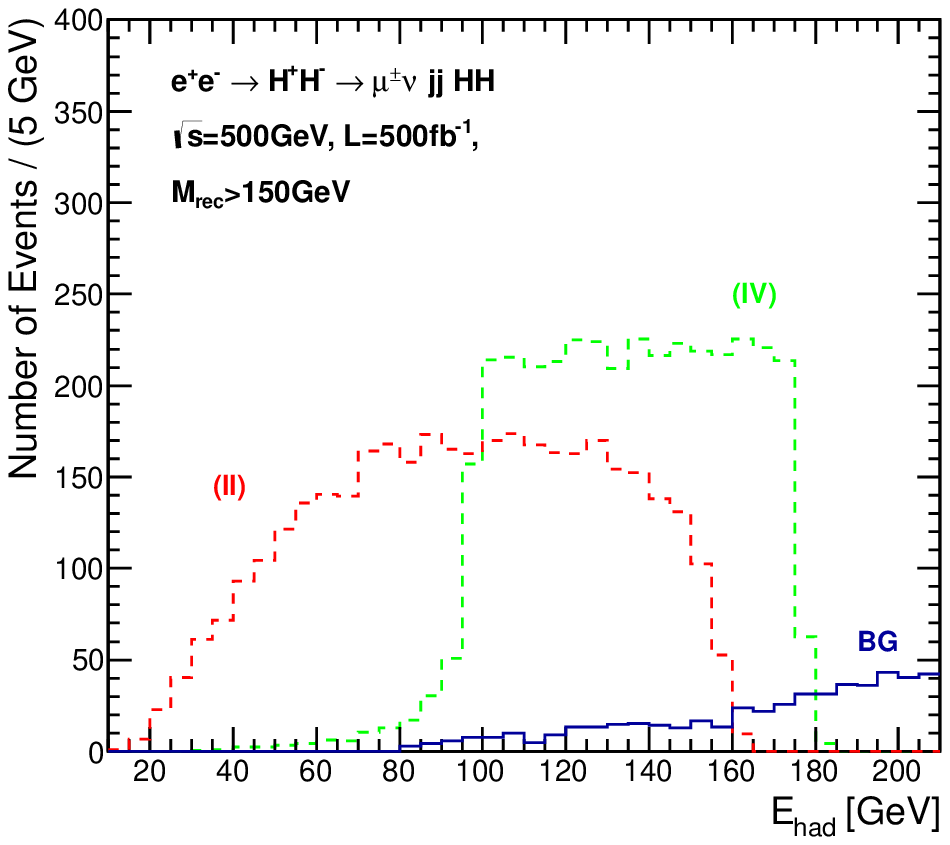}
  \caption{Distributions of $E_{\rm had}$, $M_{\rm had}$ in the
  semi-leptonic decay mode at $\sqrt{s}=250$~GeV with ${\cal L}_{\rm
  int}=250$~fb$^{-1}$ (left and middle) and
  that of $E_{\rm had}$ at $\sqrt{s}=500$~GeV with ${\cal L}_{\rm
  int}=500$~fb$^{-1}$ (right).}
  \label{fig:Semi}
 \end{center}
\end{figure}

In the left and middle panels in Fig.~\ref{fig:Semi}, distributions
of $E_{\rm had}$ and $M_{\rm had}$ in the semi-leptonic decay mode are
plotted by using the parameter set (II) at the ILC with
$\sqrt{s}=250$~GeV and ${\cal L}_{\rm int}=250$~fb$^{-1}$ with a cut of
$M_{\rm rec}>180$~GeV. 
The background contribution is negligible.

For the case with the off-shell $W$-boson, the endpoints of the all-jets
(hadrons) energy distribution are given by
\begin{align}
 E_{\rm had}^{\rm max/min} =
 \frac{\sqrt{s}}{4}\left(1-\frac{m_{H}^2}{m_{H^\pm}^2}\right)
\left[1\pm\sqrt{1-\frac{4m_{H^\pm}^2}{s}}\right]. \label{eq:1}
\end{align}
Here, we note that the invariant mass of all hadrons vanishes at the
endpoints. 
Therefore, the hadronic system would be actually observed as one jet
near the endpoints.
When we apply a cut on the smallest of the dijet invariant-mass at
$M_{\rm cut}$, the endpoints of the energy distribution would be
replaced by
\begin{align}
 E_{\rm had}^{\rm max/min} =
 \gamma_{H^{\pm}}\hat{E}_{\rm
 had}\pm\gamma_{H^{\pm}}\beta_{H^{\pm}}\,\hat{p}_{\rm had}, 
 \label{eq:4}
\end{align}
with $\gamma_{H^{\pm}} = \sqrt{s}/(2m_{H^\pm})$,
 $\beta_{H^{\pm}} = (1-4m_{H^\pm}^2/s)^{1/2}$, $\hat{E}_{\rm
 had} = (m_{H^\pm}^2-m_{H}^2+M_{\rm cut}^2)/(2m_{H^\pm})$ and
 $\hat{p}_{\rm had} = m_{H^\pm}/2\cdot\lambda(1,m_{H}^2/m_{H^\pm}^2,M_{\rm
 cut}^2/m_{H^\pm}^2)$. 
Thus, the mass information can be still obtained. 
Furthermore, the maximum value of the invariant mass of all hadrons is
just the difference between $m_{H^\pm}$ and $m_{H}$, 
\begin{align}
M_{\rm had}^{\rm max}=m_{H^\pm}-m_{H}. \label{eq:2}
\end{align}

In the right panel in Fig.~\ref{fig:Semi}, the $E_{\rm had}$
distribution in the semi-leptonic decay modes are plotted by using the
parameter sets (II) and (IV) at the ILC with $\sqrt{s}=500$~GeV and
${\cal L}_{\rm int}=500$~fb$^{-1}$ with a cut of $M_{\rm rec}>150$~GeV. 
Notice that the parameter set (II) corresponds to the case where
$H^{\pm}$ decays into off-shell $W$ and $H$, and (IV) corresponds to the
case where $H^{\pm}$ decays into on-shell $W$ and $H$.
When the $W$-boson is on-shell, the signal distribution is like a
rectangle where the edges are given by $E^{\rm max/min}_{\rm had}$ in 
Eq.~(\ref{eq:4}), but with $M_{\rm cut}$ being replaced by $m_{W}$. 

We note that the dijet system in the semi-leptonic decay mode may be
replaced by the dijet subsystem which satisfies $M_{jj}\simeq m_W$ in
the all-hadronic decay mode where the signature is 
four jets plus missing energy (see below).

Now, we turn to the all-hadronic decay mode, which results
into the signature of four jets plus large missing energy.
Major SM background comes from the production of four partons and two
neutrinos. 
\begin{figure}[t]
 \begin{center}
  \includegraphics[width=0.328\textwidth,clip]{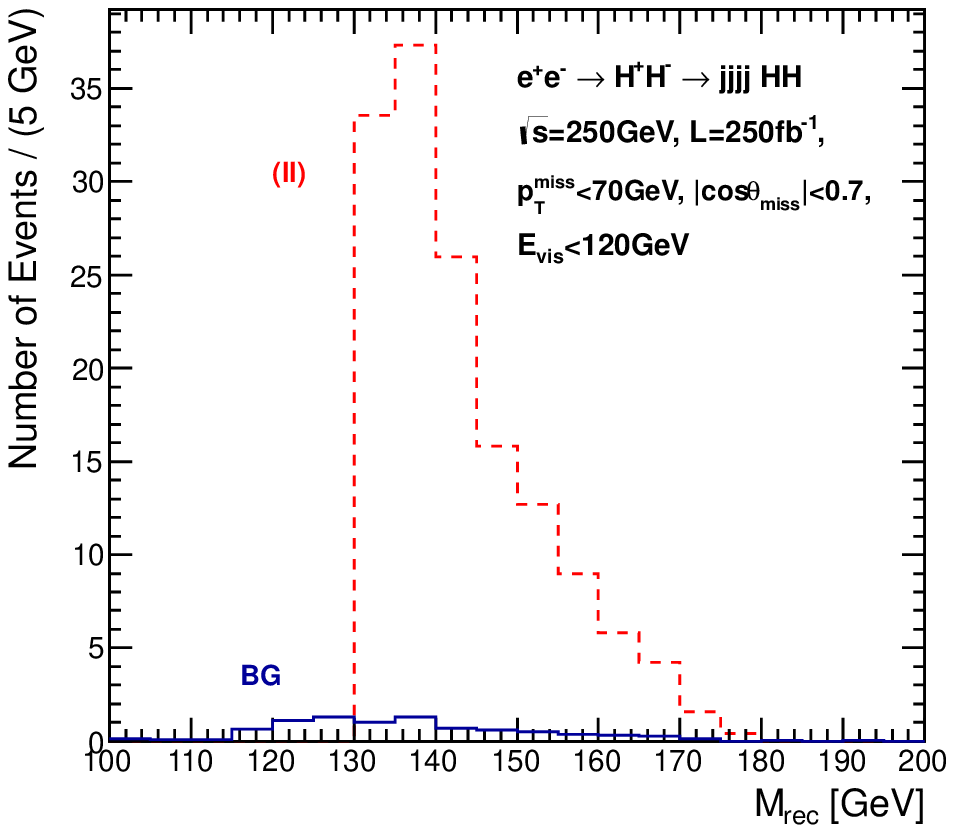}
  \includegraphics[width=0.328\textwidth,clip]{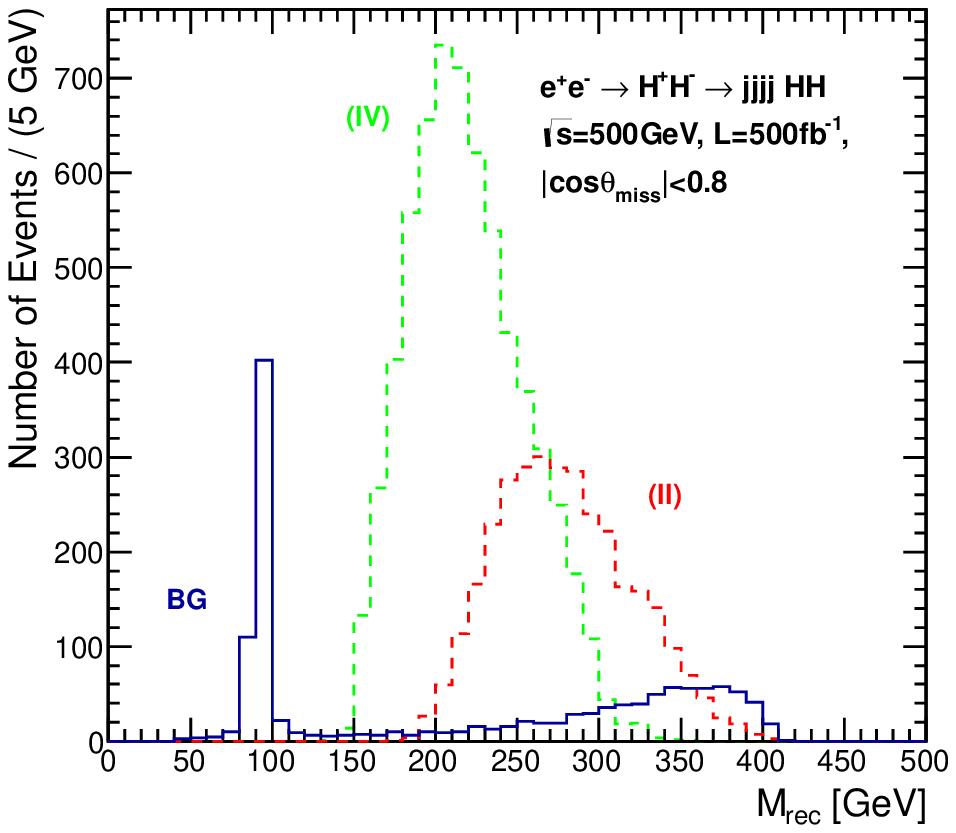}
  \includegraphics[width=0.328\textwidth,clip]{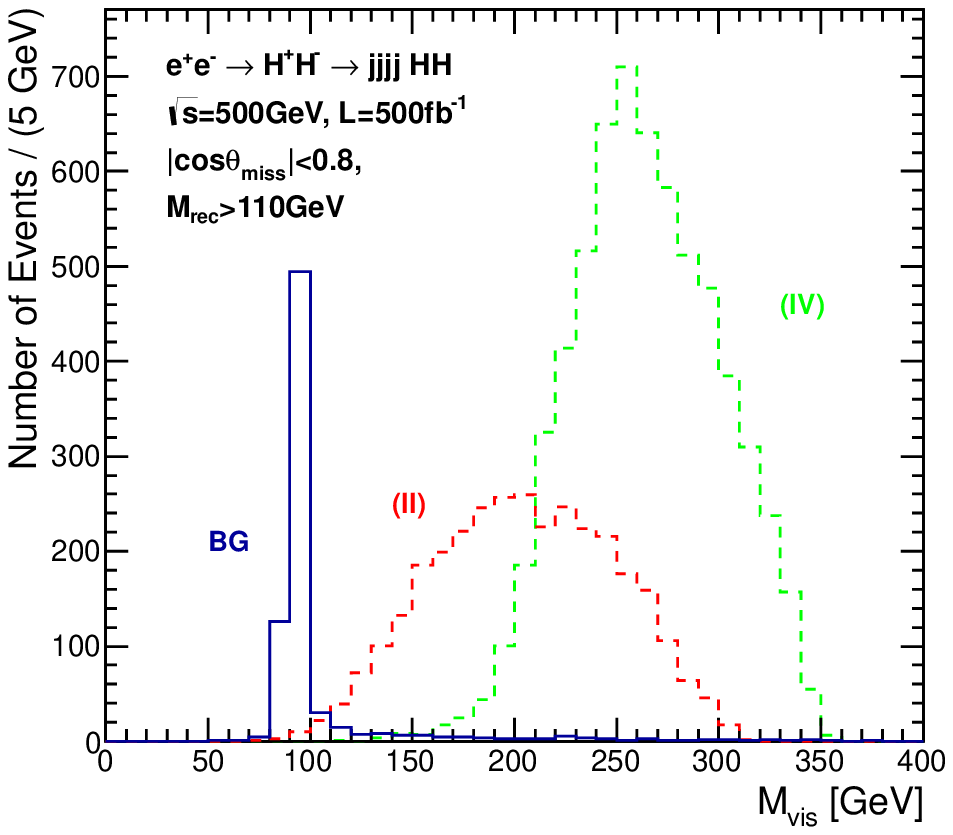}
  \caption{Distributions of $M_{\rm rec}$ in the all-hadronic decay
  mode at $\sqrt{s}=250$~GeV with ${\cal L_{\rm
  int}}=250$~fb$^{-1}$ (left), and $M_{\rm rec}$ and $M_{\rm 
  vis}$ distributions in the all-hadronic mode at $\sqrt{s}=500$~GeV
  with ${\cal L_{\rm int}}=500$~fb$^{-1}$ (middle and right).}
  \label{fig:Dist_On}
 \end{center}
\end{figure}
In the left panel of Fig.~\ref{fig:Dist_On}, $M_{\rm rec}$ distribution
is plotted for the signal process using the parameter set (II) at
$\sqrt{s}=250$~GeV with ${\cal L}_{\rm int}=250$~fb$^{-1}$. 
To reduce the SM background, kinematical cuts of $p_{T}^{\rm
miss}>70$~GeV, $|\cos\theta_{\rm miss}|<0.7$ and $E_{\rm vis}<120$~GeV
are applied, where $\theta_{\rm miss}$ is the polar angle of the missing
3-momenta and $E_{\rm vis}$ is the sum of the energy of all hadrons in
one event. 
As a result, the SM background is sufficiently reduced. 
The minimum of the $M_{\rm rec}$ distribution is at the twice of $m_H$, 
\begin{align}
 M_{\rm rec}^{\rm min}=2m_{H}.\label{eq:recmin}
\end{align}

In the middle panel of Fig.~\ref{fig:Dist_On}, the same distributions
are plotted but for the signal processes using parameter sets (II) and
(IV) at $\sqrt{s}=500$~GeV with ${\cal L_{\rm int}}=500$~fb$^{-1}$.
By the kinematical cut of $|\cos\theta_{\rm miss}|<0.8$, the SM
background is sufficiently reduced except at $M_{\rm rec}\simeq m_{Z}$.
As it is shown in Appendix, the peak of the signal distribution
is given by 
\begin{align}
 M_{\rm rec}^{\rm peak} = \frac{m_{H}\sqrt{s}}{m_{H^{\pm}}}.
\label{eq:recpeak}
\end{align}
It is the advantage of this observable that this relation holds even
when the $W$-boson in $H^\pm\to W^\pm H$ is off-shell.
Thus, the ratio of $m_{H}$ and $m_{H^{\pm}}$ can be determined.

In the right panel of Fig.~\ref{fig:Dist_On}, the $M_{\rm vis}$
distributions are plotted for the signal processes using parameter
sets (II) and (IV) at $\sqrt{s}=500$~GeV with ${\cal L_{\rm
int}}=500$~fb$^{-1}$. 
In addition to the kinematical cut applied in the previous panel,
the cut of $M_{\rm rec}>110$~GeV is applied to reduce the SM background
which has a missing energy from $Z\to\nu\bar\nu$. 
After these cuts, the SM background is sufficiently reduced except at
$M_{\rm vis}\simeq m_Z$. 
The signal distribution has a peak at 
\begin{align}
 M_{\rm vis}^{\rm peak}=\frac{m_{W}\sqrt{s}}{m_{H^{\pm}}},
\label{eq:vispeak}
\end{align}
when the $W$-boson in $H^\pm\to W^\pm H$ is on-shell [the case
(IV)]. 
When the $W$-boson is off-shell, the relation on the peak position no
more holds.

\subsection{Mass Determination}

\begin{figure}[t]
 \begin{center}
  \includegraphics[width=0.49\textwidth,clip]{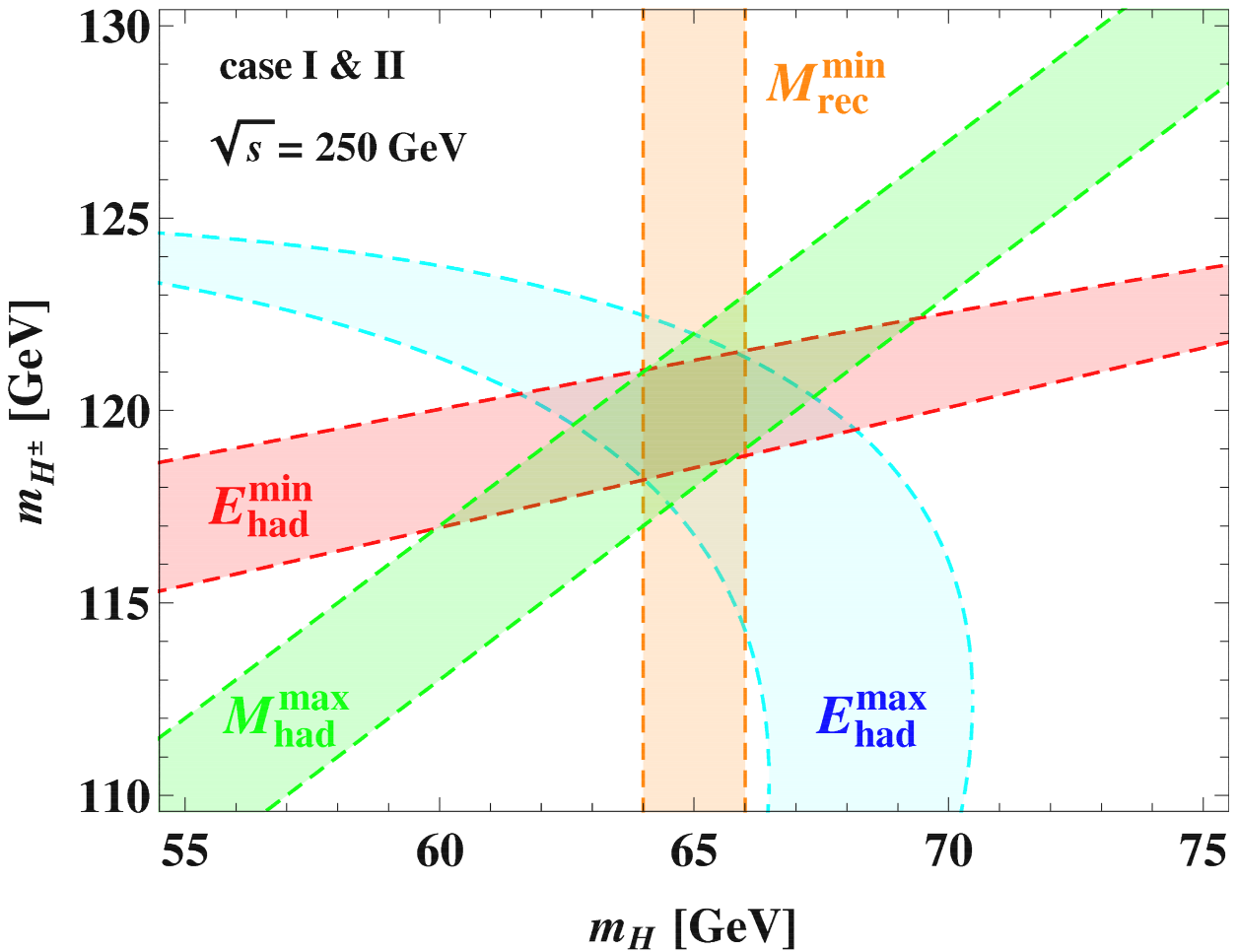}
  \includegraphics[width=0.49\textwidth,clip]{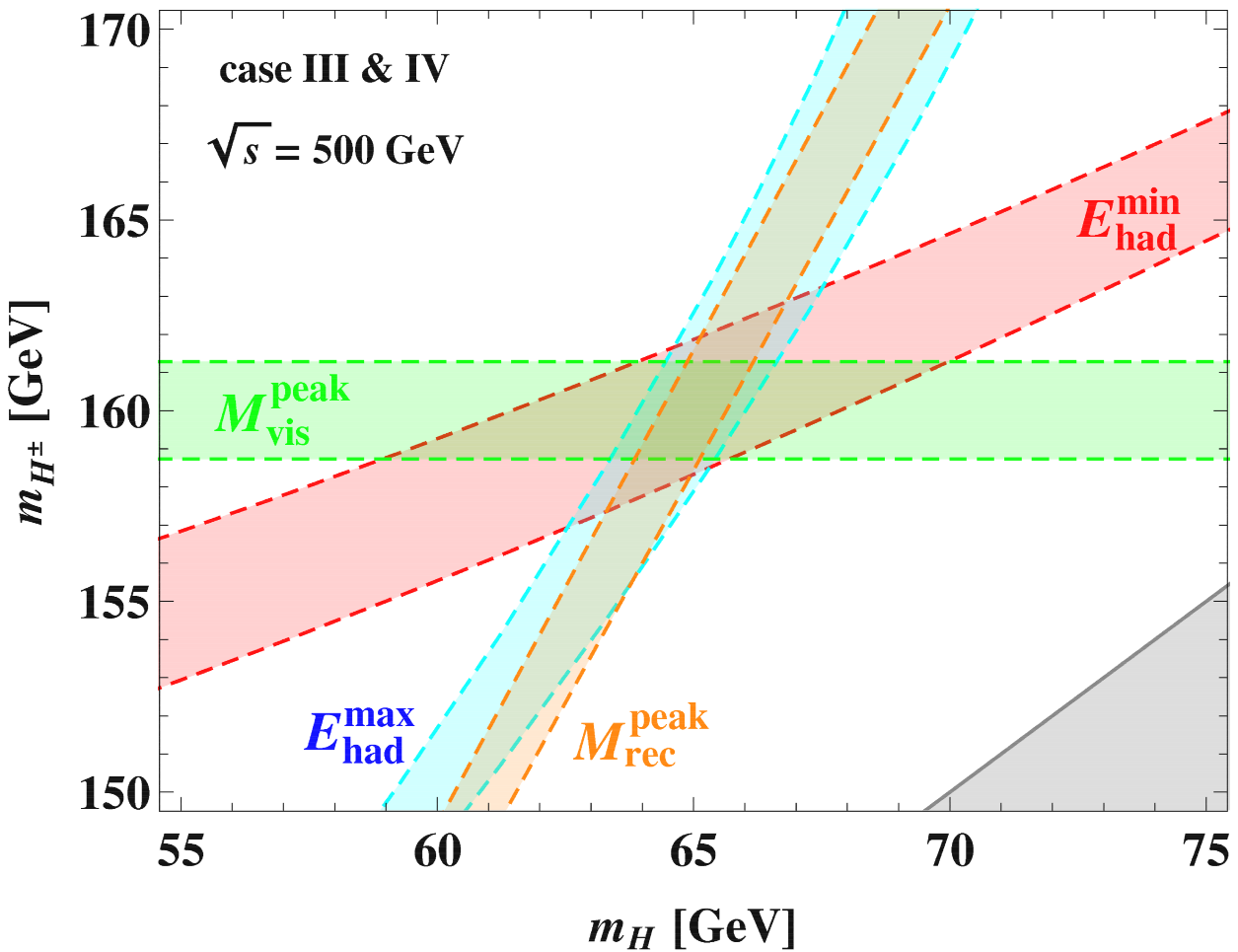}
  \caption{Determinations of $m_{H^\pm}$ and $m_{H}$ by the four
  observables are illustrated in the left [right] panel for the cases
  (I, II) [(III, IV)] at $\sqrt{s}=250$~GeV [500~GeV].
  Each observable is assumed to be measured in $\pm 2$~GeV accuracy.
  } \label{fig:Mass}
 \end{center}
\end{figure}

Here, we summarize the observables for determining the masses of inert
scalars. 
First, we consider the determination of $m_{H^{\pm}}$ and $m_{H}$ in the
process $e^+e^-\to H^+H^-$.
If $m_{H^{\pm}}-m_{H}<m_{W}$, $m_{H^\pm}$ and $m_{H}$ can be determined
simultaneously by measuring the four quantities; $E^{\rm max}_{\rm
had}$, $E^{\rm min}_{\rm had}$ in Eq.~(\ref{eq:1}), $M^{\rm max}_{\rm
had}$ in Eq.~(\ref{eq:2}) and $M^{\rm min}_{\rm rec}$ in
Eq.~(\ref{eq:recmin}).
In the left panel of Fig.~\ref{fig:Mass}, we show how the masses are
determined by the measurements of the four quantities for the cases (I)
and (II) at $\sqrt{s}=250$~GeV.
The four bands are plotted on the $m_{H^\pm}$-$m_{H}$ plane by
assuming that the four quantities are measured in $\pm 2$~GeV accuracy
(without any systematic shifts).
For this assumption, the accuracy of the $m_{H^\pm}$ ($m_{H}$)
determination would be $\pm 2$~GeV ($\pm 1$~GeV).
We note that our assumption is a simplest example in order to clarify
the relation between the accuracy of the mass determination and that of
the observables without involving experimental details.
If the accuracy of the observable is changed, the width of the band on
the figure varies in proportion to that.
On the other hand, if $m_{H^{\pm}}-m_{H}\ge m_{W}$, the four
observables, $E_{\rm had}^{\rm max}$, $E_{\rm had}^{\rm min}$ in
Eq.~(\ref{eq:4}) but with replacing $M_{\rm cut}$ by $m_W$, $M_{\rm
rec}^{\rm peak}$ in Eq.~(\ref{eq:recpeak}) and $M_{\rm vis}^{\rm peak}$
in Eq.~(\ref{eq:vispeak}) are utilized for the mass determination. 
In the right panel of Fig.~\ref{fig:Mass}, the four bands are plotted on
the $m_{H^\pm}$-$m_{H}$ plane by assuming that the four
observables are measured in $\pm 2$~GeV accuracy. 
It turns out that the constraints from measurements of $M^{\rm
peak}_{\rm vis}$ and $M^{\rm peak}_{\rm rec}$ are more stringent than
those from the $E^{\rm max/min}_{\rm had}$ measurements, if these
quantities are measured in an equal accuracy. 
It is expected that peak positions can be precisely determined more than
endpoints of distributions in the presence of the resolution of energy
measurements and the remaining background contributions. 
By combining the four measurements with the uncertainty of $\pm 2$~GeV,
$m_{H^\pm}$ and $m_{H}$ can be determined in $\pm 1$~GeV accuracy. 

Next, the determination of $m_{A}$ can be achieved by combining the
observables in the process $e^+e^-\to HA$.
For the cases with the off-shell $Z$-boson (I, II, III),
$E_{\ell\ell(jj)}^{\rm max}$ in Eq.~(\ref{eq:endoff}) and $M^{\rm
max}_{\ell\ell(jj)}$ in Eq.~(\ref{eq:mmax}) measurements can be
utilized. 
However, at $\sqrt{s}=250$~GeV and $\sqrt{s}=500$~GeV, since the two
constraints are very similar, these masses cannot be determined at one
point.
In that case, one needs the value of $m_{H}$ fixed in the process
$e^+e^-\to H^+H^-$ as an input to determine $m_A$.
While for the case with the on-shell $Z$-boson (IV),
measurements of $E_{\ell\ell(jj)}^{\rm max/min}$ in Eq.~(\ref{eq:endon})
can be utilized. 
In that case, the expected accuracy of the mass determination is
$\pm3$~GeV for the measurement of the observables in $\pm2$~GeV
accuracy. 

\section{Discussions}\label{sec:dis}

At the ILC, another important measurement which we have not discussed is
the spin measurement of the extra particles. 
In contrast to $H^\pm$ and $(H,A)$ in the IDM, supersymmetric model
contains charginos and neutralinos which are fermions, and the littlest
Higgs model with $T$-parity or models with extra-dimension scenario
predict heavy $W$-bosons and heavy photons which are vector particles. 
Since the production and decay of these particles can mimic the
signatures studied in this letter, it is particularly crucial to
discriminate models by the spin measurement of new particles. 
The spin of new particles could be traced in the threshold behavior of
the production cross sections, the scattering angular dependence, and
the decay angular dependence.
It has been discussed that these measurements are possible for the
processses we considered at the
ILC~\cite{Choi:2006mr,Asano:2011aj,Ginzburg:2012zr}. 
On the contrary, those measurements at the LHC are difficult, because
the center-of-mass energy in the partonic scattering is not fixed and
also the center-of-mass system of the partonic scattering cannot be
reconstructed.

Finally, we comment on the total comparison for the IDM studies at the
LHC and the ILC.
It is shown that the LHC has discovery potential for the inert scalars
by using the signatures with multileptons plus large missing transverse
momentum~\cite{Dolle:2009ft,Miao:2010rg,Gustafsson:2012aj}, if the mass
difference is sufficiently large. 
Thus, in some preferable situations, the evidence of the inert scalars
would show up at the LHC before the ILC experiment starts. 
On the other hand, as we have shown, the ILC can probe the case with
smaller mass difference still in a good accuracy, as long as the
production processes are kinematically accessible, i.e.\
$\sqrt{s}>2m_{H^{\pm}}$ and $m_{H}+m_{A}$.
Thus, there is a parameter region where any evidence can not be observed
at the LHC, but the ILC could find it.
In any case, at the ILC, the masses of the inert scalars can be
reconstructed in good accuracy, and furthermore, the spin of the
particles can be measured in principal.
Therefore, the final discrimination of the model beyond the SM would be
performed at the ILC. 

\section{Conclusions}\label{sec:conc}

To conclude, we have studied the collider phenomenology of the inert
scalar particles at the ILC in the IDM. 
The model contains a scalar dark matter candidate which is stable due to
the unbroken $Z_2$ symmetry, in addition to another neutral scalar and 
charged scalar bosons. 
At collider experiments, because the dark matter would escape from
the detector, the signatures always include large missing energy.
We have studied the collider signatures of the pair production of the
charged scalars and the associated production of the two neutral
scalars with appropriate kinematical cuts to reduce the SM backgrounds.
We have also investigated the observables to determine the masses of the
scalars in these processes. 
We have shown that distinctive signatures can be observed even if the
cases with small mass difference by applying simple kinematical cuts,
and that by combining these observables the inert scalar masses can be
reconstructed in a good accuracy at the ILC.

\acknowledgments
M.A.\ thanks L.~Lopez Honorez for useful discussions.
S.K.\ and H.Y.\ would like to thank Keisuke Fujii for useful discussions.
The paper was supported in part by Grant-in-Aid for Scientific Research,
No.\ 24340036.
M.A.\ was also supported in part by Grant-in-Aid for Scientific
Research, No.\ 22740137.
S.K.\ was also supported in part by Grant-in-Aid for Scientific
Research, Nos.\ 22244031 and 23104006.

\renewcommand{\theequation}{A\arabic{equation}}
\setcounter{equation}{0}
\section*{Appendix: Invariant-mass distributions in $X\to AA\to BCBC$}

Here, we consider the kinematics of a general process $X\to A_1A_2\to
B_1C_1B_2C_2$, where $X$ is a certain initial-state with the fixed
collision energy, $A_i$ are scalars with the same mass $m_A$, e.g.\ the
identical particles or the charge-conjugate particles.
$B_i$ and $C_i$ are any particles with their masses $m_B$ and $m_C$,
respectively, which are produced from the isotropic decay of $A_i$. 
For the meantime, we consider the case where all the particles are
on-shell, i.e.\ $\sqrt{s}\ge 2m_{A}\ge2m_{B}+2m_{C}$. 
We consider the invariant-mass distribution of the $B_1B_2$ pair: 
\begin{align}
 \frac{d\sigma}{dM_{BB}^2} \propto & \int d\Phi_4(X;B_1C_1B_2C_2)\cdot
 \delta\left(M_{BB}^2-(p_{B_1}+p_{B_2})^2\right)
 \nonumber\\
 & \times\frac{1}{[(p_{B_1}+p_{C_1})^2-m_A^2]^2+m^2_A\Gamma^2_A}\cdot
 \frac{1}{[(p_{B_2}+p_{C_2})^2-m_A^2]^2+m^2_A\Gamma^2_A},
\end{align}
where $d\Phi_n$ is the $n$-body phase-space volume element and $\Gamma_A$
is the total decay width of $A$.
Using the narrow width approximation for $A$, it is calculated as 
\begin{align}
 \frac{d\sigma}{dM_{BB}^2} \propto & \int d\Phi_2(A_1;B_1C_1)
 \,d\Phi_2(A_2;B_2C_2)
 \,\delta\left(M_{BB}^2-(p_{B_1}+p_{B_2})^2\right)\nonumber\\
 \propto & \int d\cos\theta_1 d\cos\theta_2 d\phi\,
 \delta{\left(M_{BB}^2-(p_{B_1}+p_{B_2})^2\right)},\label{eq:int}
\end{align}
where $(\theta_{1}$, $\phi)$ and $\theta_2$ are the decay angles in the 
rest-frame of $A_{1}$ and $A_{2}$, respectively.
The azimuthal angle in the decay of $A_2$ is fixed to be zero, and the
scattering angles in the process $X\to A_1A_2$, which are irrelevant to 
$M_{BB}$, are integrated out.
By using the above integration variables, $(p_{B_1}+p_{B_2})^2$ is
expressed as 
\begin{align}
 (p_{B_1}+p_{B_2})^2 &= \nonumber\\
 2m_B^2 &+\frac{s}{8}\left[
 \left(1+\beta_A^2\right)(\epsilon^2+\beta^2c_1c_2)
 +2\beta_A\,\epsilon\beta(c_1+c_2) 
 +(1-\beta_A^2)\beta^2s_1s_2c_\phi\right]. \label{eq:Mbb}
\end{align}
Here, $s=p_X^2$, $\beta_{A}=\sqrt{1-4m_{A}^2/s}$, $\epsilon =
1+m_B^2/m_A^2-m_C^2/m_A^2$, $\beta =
\lambda(1,m_B^2/m_A^2,m_C^2/m_A^2)$ with
$\lambda(a,b,c)=\sqrt{a^2+b^2+c^2-2(ab+bc+ca)}$, $c_i=\cos\theta_i$ for
$i=1,2$ and $c_\phi=\cos\phi$. 
Although the numerical integration in Eq.~(\ref{eq:int}) is
straightforward, we find that, unless $\sqrt{s}\simeq2m_A$, it is a good
approximation to neglect the last term in Eq.~(\ref{eq:Mbb}). 
In that case, the $M_{BB}$ distribution can be analytically expressed as 
\begin{subequations}
\begin{align}
 \frac{d\sigma}{dM_{BB}^2}\propto
 \log\left[
 \frac{4\left\{(s-2m_A^2)(M_{BB}^2-2m_B^2)-m_A^4\epsilon^2\right\}}
 {s^2\left\{(1+\beta_A^2)\beta-2\beta_A\epsilon\right\}}\right],
\end{align}
for $M_{\rm min}\leq M_{BB}\leq m_{B}\sqrt{s}/m_A$, and 
\begin{align}
 \frac{d\sigma}{dM_{BB}^2}\propto
 \log\left[
 \frac{s^2\left\{(1+\beta_A^2)\beta+2\beta_A\epsilon\right\}}
 {4\left\{(s-2m_A^2)(M_{BB}^2-2m_B^2)-m_A^4\epsilon^2\right\}}\right],
\end{align} \label{eq:dist}
\end{subequations}
for $m_{B}\sqrt{s}/m_A<M_{BB}\leq M_{\rm max}$, where
$M_{\rm max/min}=\sqrt{s}/2\cdot(\epsilon\pm\beta_A\beta)$.
Thus, we find that the distribution has a peak at $m_B\sqrt{s}/m_A$, 
which therefore depends only on the ratio $m_B/m_A$ but not on $m_C$.

Here we give some comments.
In case with $m_{A}\le m_{B}+m_{C}$, if the particles $C_i$ are 
off-shell, the peak position of the $M_{BB}$ distribution is still given
by $\sqrt{s}\,m_B/m_A$.
On the other hand, if the particles $B_i$ are off-shell, the
reconstructed $M_{BB}$ distribution is largely deformed. 
The $M_{CC}$ distribution can be obtained by replacing
$B\leftrightarrow C$ from the $M_{BB}$ distribution in
Eqs.~(\ref{eq:dist}). 
Thus, the peak position of the $M_{CC}$ distribution is given by $M^{\rm
peak}_{CC}=m_C\sqrt{s}/m_A$, independently from $m_B$. 

At $e^+e^-$ colliders where the total 4-momenta of the
collision can be fixed, it is possible to assess the invariant mass of
the all missing particles (the recoil mass) by $M_{\rm
rec}^2=(p_{\rm in}-p_{\rm vis})^2$, where $p_{\rm in}$ is the total
4-momenta of the initial $e^+e^-$ system.
Therefore, the invariant mass of the pair of two missing particles can
be reconstructed, if these are the only missing particles in the event. 



\end{document}